\documentclass[twocolumn,aps,prl,showpacs,preprintnumbers,amsmath,amssymb]{revtex4}
\usepackage{graphicx}
\usepackage{bm}
\usepackage{gensymb}
\usepackage{color}
\begin{document}
\title{A scaling theory of quantum breakdown in solids}
\author{Bikas K. Chakrabarti and Debashis Samanta\footnote{Corresponding author: debashis.samanta@saha.ac.in}}

\affiliation{ Theoretical Condensed Matter Physics Division,
  Saha Institute of Nuclear Physics, 1/AF Bidhannagar, Kolkata 700 064, India}
\date{\today}

\begin{abstract}
We propose a new scaling theory for general quantum breakdown phenomena. 
 We show, taking  
 Landau-Zener type breakdown as a particular example, that the breakdown 
 phenomena can be viewed as a quantum phase transition for which the 
 scaling theory is developed. The application of this new scaling theory
 to Zener type breakdown in Anderson insulators, and quantum quenching has 
 been discussed.
\end{abstract}
\maketitle

The behavior of solid state electronic systems in non-equilibrium conditions,
 particularly the insulator-metal transitions leading to dielectric
 breakdown in constant electric field just beyond a particular threshold, 
 has been a matter of considerable investigation.
 This started with Zener \cite{Zener:1932,Zener:1934}, 
 considering the charge-excited carrier 
 generation process as tunneling through the band gap $\triangle$, from 
 ground state to excited states in one-dimensional band insulator, using the 
 time-independent as well as time-dependent 
 gauge \cite{Zener:1934,Landau:1932}.
 Recent years have noticed increasing interests in the nonlinear phenomena
 of quantum breakdown in many-body systems put in strong dc electric fields, 
 in 
 intense laser fields, etc. The interest also comes because  of the fact that 
 the high sensitivity 
 (drastic changes) of electronic states to external field near the phase 
 boundary finds its application in dielectric breakdown based 
 current-switching   
 devices \cite{Asamitsuetal:1997}. However, more interesting is its 
 relevance to non-equilibrium statistical physics where one can observe the 
 behavior of uncorrelated as well as strongly correlated electronic system
 near phase boundary under non-equilibrium conditions and in the study of 
 quantum defect production in rapid quench of quantum fluctuations 
 (see~e.g., \cite{DasCha:2008}).

The tunneling rate $P$ of charge excited carriers through a gap $\triangle$ 
 in dc electric field $E$ (for the case of band insulator) has been shown to 
 be expressed in closed form,
 which is
 proportional to $\exp{\left[-(\mathcal{E}_{th}/Ee)\right]}$; with
 $ \mathcal{E}_{th}\sim {\triangle}^2$ and $e$ denoting the carrier charge.
 Recently, it has been shown that such Landau-Zener transition estimate can 
 be straightforwardly extended to Mott insulators as 
 well with the same exponent value \cite{OkaAoki:2009,Okaetal:2003}, 
 though the carriers are different. In band 
 insulator holes and electrons undergo tunneling through band gap, whereas
 in Mott insulator doublons and holes carry out the same role through Mott
 gap. 

Within the present framework, it seems to be virtually impossible to obtain 
 such an exact expression for the tunneling rate of the 
 carriers in a disordered 
 quantum systems like the Anderson insulators \cite{LeeRam:1985}. Our 
 purpose here is to 
 develop a 
 simple scaling theory for such quantum breakdown phenomenon and explore the
 Landau-Zener type \cite{Landau:1932} breakdown in, e.g., 
 Anderson insulators (say, in three 
 dimension). It may be mentioned 
 that the growth of fluctuation correlations near various well-studied 
 classical breakdown points have already helped establishing a phase 
 transition picture, leading to well established scaling theories for 
 phenomena, like fracture etc. 
 (see e.g., \cite{ChaBen:1997}). Also, 
 precise 
 critical behavior of breaking 
 in, e.g., the fiber bundle like models of fracture are now quite 
 established 
 \cite{Pradhanetal:2009}.

From elementary quantum mechanics (see e.g., \cite{Marz:1961}), 
 one gets the tunneling rate
 through an energy barrier of (large) width $w$ to be proportional 
 to $\exp(-c\kappa w)$, where $\kappa$ is the damping factor or `imaginary 
 momentum' vector determined by the height of the barrier and $c$ is some 
 dimensionless constant. Considering the 
 quantum breakdown phenomena as 
 quantum phase transitions (see e.g., \cite{Sachdev:1999}), we assume that 
 near the breakdown point, the 
 macroscopic correlation length $\xi$ will provide the only scale governing the behavior 
 of  
 physical quantities of the system. We assume this to be true for the 
 tunneling rate
 $P$ as well. Hence, the width $w$ is scaled by $\xi$ and 
 `momentum' is considered to be proportional to inverse of $\xi$. Hence, the 
 dimensionless tunneling rate expression takes the form 
\begin{eqnarray}
\label{eqn:P}
   P  & \sim & \exp\left[-c\left (\frac{a}{\xi}\right )
                    \left (\frac{w}{\xi}\right )\right].
\end{eqnarray}
 Here, $a$ denotes lattice constant, the only other (microscopic) length scale
 in the system. The above gives our proposed scaling form for tunneling 
 probability across a barrier of width $w$ near a quantum phase transition 
 characterised by the correlation 
 length $\xi$ determined by the barrier height. 

We now proceed with the application of the above scaling ansatz.
 In presence of a constant dc field, the bands effectively get tilted in 
 field direction, which causes each energy band degenerate with the others. 
 Therefore, electron   can pass from lower energy band to upper one 
 traversing an effective 
 distance $w$ in the direction of field and we envisage the process involved 
 as tunneling.
 The effective width $w$ can be estimated
 assuming that the energy $Eew$ acquired from the field $E$ over the width $w$
 will compare with the 
 gap $\triangle$, giving $w=\triangle/Ee$. 
  We assume,  
 the correlation length diverges as $\triangle^{-\nu}$ at 
 criticality $(\triangle = 0)$ with exponent $\nu$ \cite{Sachdev:1999}, and 
 additionally we assume that even up to the breakdown point the external field 
 does not affect the correlation length exponent.  
 Therefore the tunneling rate expression becomes
\begin{eqnarray}
\label{eqn:mp}
  P  \sim  \exp\left[-\left(\frac{ca}{Ee}\right){\triangle}^{\gamma}\right];
            \hspace{1mm} \gamma=1+2\nu .
\end{eqnarray}

For the band and Mott insulators, putting $\nu$ as $1/2$ 
 (see e.g., \cite{Imada_etal:1998}), 
 we get $P \sim \exp(-\mathcal{E}_{th}/Ee)$, 
 $\mathcal{E}_{th} \sim \triangle^2$, the 
 same expression as was obtained 
 earlier \cite{Zener:1934,Landau:1932,OkaAoki:2009,Okaetal:2003}. For 
 Mott insulators, the 
 recent 
 study by Oka {\it{et al.}} \cite{OkaAoki:2009,Okaetal:2003} supports the 
 above expression as well, where the
 gap is determined by the electronic correlation. 

Since the system response near breakdown is governed essentially by the 
 correlation length
 exponent, one can make some estimate for the electrical response in quantum 
 systems with disorder, equipped with the knowledge of its correlation length 
 around 
 criticality in such disordered systems.  
 Studies on Anderson transition show that the electron as a quantum particle 
 can not  diffuse through the geometrically 
 percolating path due to the coherent
 back scattering (of the wave function) from the random geometry of the 
 clusters in dimensions less than three \cite{LeeRam:1985}. Since 
 all 
 the states on any such percolating lattice gets 
 localized (exponentially), electrons do not diffuse through the 
 disordered (classically percolating) lattice. 
 In three dimension (for 
 concentrations above the quantum percolation threshold), if 
 the Fermi level $\epsilon_f$ lies below the mobility 
 edge $\epsilon_c$ $(\not=0$ for three dimension$)$, the system
 remains to be quantum mechanically non-percolating with the exponentially
 localized electronic states, while 
 for $\epsilon_f>\epsilon_c$ the states are extended. Hence 
 if $\epsilon_f<\epsilon_c$, the system is an insulator and 
 for $\epsilon_f>\epsilon_c$ it is a conductor. The scaling property in 
 this conducting region is well established; in 
 particular the localization or correlation 
 length $\xi \sim |\epsilon_f-\epsilon_c|^{-\nu}$ and precise estimates 
 (both theoretical and experimental) of $\epsilon_c$ and $\nu$ are now 
 available (see e.g. \cite{LeeRam:1985,Belitz}).   
 For $\epsilon_f<\epsilon_c$, one can think of an insulator-metal
 transition in such a non-interacting random system, analogous to quantum 
 breakdown in band or Mott insulators, induced by (strong) dc electric 
 fields. The critical behavior can be 
 easily predicted utilizing the precise knowledge of localization length 
 exponent $\nu$. With $\nu \simeq 1$ \cite{LeeRam:1985} for 
 Anderson 
 insulators in three dimension, the tunneling probability across the 
 mobility gap $\triangle \equiv |\epsilon_c-\epsilon_f|$ near breakdown point, 
 will be given by 
 $P \sim \exp(-\mathcal{E}_{th}/Ee)$, where 
 $\mathcal{E}_{th} \sim \triangle^{\gamma}$; $\gamma \simeq 3$. This
 indicates that $\ln P$ would scale  as $\triangle^3$ for
 Zener type breakdown in Anderson insulators, instead of as $\triangle^2$ for
 similar breakdown in band insulators. 
     
In the context of adiabatic quantum computations, where one exploits the 
 tunneling probability through the energy or cost barriers for local minima
 to reach a global minimum (may be degenerate, as in spin glass), one needs
 to estimate the density of defects remaining over the ground state as one 
 quenches from a highly excited (para) state \cite{DasCha:2008}. For
 example, in a transverse Ising glass model represented by the hamiltonian
 \cite{DasCha:2008} 
\begin{equation}
 H=-\sum_{(i,j)}^{N} J_{ij} \sigma_i^z \sigma_j^z - \Gamma(t) \sum \sigma_i^x ,
\end{equation}
 with random exchanges $J_{ij}$ between the Pauli spins 
 $\vec{\sigma}$,
 one can vary the quantum fluctuation slowly but linearly 
 $(\Gamma(t)=1-(t/\tau)$;
 $0 \leq t \leq \tau)$ to arrive at a ground state of the classical spin glass
 $(\mbox{for } \Gamma = 0 \mbox{ at } t=\tau)$  from a disordered 
 or a para phase  $(\mbox{for } \Gamma =1 \mbox{ at } t=0)$. The amount of 
 defect over the true 
 ground state (giving the solution of a 
 computationally hard problem) depends on the quenching time $\tau $  
 and the 
 advantage of such 
 annealing or quenching (tunneling) through macroscopically high or $O(N)$ 
 barriers (instead of thermal hopping over the barriers as in classical 
 simulated annealing \cite{Kirkpatrick_etal:1983}) might be lost if defect 
 concentration remains high. Identifying the force $Ee$ in (\ref{eqn:mp}) as 
 the rate $R$ of change in the energy of the system, one 
 can easily estimate the density of defects by calculating the tunneling 
 probability as 
\begin{eqnarray}
\label{eqn:prob}
 P  \sim  \exp \left (-\frac{\triangle^{\gamma}}{R}\right )
    \sim  \exp (-\tau \triangle^{1+2\nu}),
\end{eqnarray}
  where $R \sim \frac{\partial}{\partial t}<H(t)>$.
  Assuming that the residual energy $E_{\mathit{res}}$ (over the ground state 
  energy) of the
  quenched state to be proportional to 
  $\int \! \! \triangle \, P \rho(\triangle)\, d\triangle$, and averaging over 
  the Gaussian distribution
  $\rho(\triangle) \sim \exp(-\triangle^{2})$\cite{BindYoung:1986} 
  of the local gap parameter
  $\triangle$ or of the local fields $(h_{\mathit{loc}};\hspace{1mm}
  \triangle=\sqrt{h_{\mathit{loc}}^2+\Gamma^{2}})$\cite{DasCha:2008}, one
  gets the decay rate of average residual energy as 
  $E_{\mathit{res}} \sim \tau^{-\frac{4}{3}}$  
  (for large $\tau$ values) in a 
  long-range Gaussian spin glass model (having $\nu=1/4$ \cite{Readetal:1995}).
  This, in fact, 
  compares
  quite well with the numerical estimate of $E_{\mathit{res}}$ for fairly 
  larger Gaussian  spin glass systems studied in \cite{Matsudaetal:2009}, 
  where $E_\mathit{res}$
  decays much slower than $\tau^{-2}$ (observed for smaller systems).   

In brief, we propose here a generalized scaling form (\ref{eqn:P}) for the 
 single particle
 tunneling probability across a barrier, where we  subsequently replace the single particle 
 correlation by the many body one appropriate for a quantum many body phase 
 transition, obtaining thereby 
 the generalized Landau-Zener breakdown probability. This scaling form has 
 then been utilized for a Zener type breakdown in Anderson insulators in three
 dimension. Finally,
 we apply it to estimate the quenching rate dependence of the residual 
 energy of a long range quantum Ising spin glass when its transverse field
 is swept through the critical point. Comparision with numerical results seems
 encouraging.

We are thankful to Arnab Das, Jon Eriksen, Pradeep K. Mohanty and Purusattam 
 Ray for important 
 discussions.

\end{document}